# THEORETICAL CALCULATIONS FOR PREDICTED STATES OF HEAVY QUARKONIUM VIA NON-RELATIVISTIC FRAME WORK

A. M. Yasser[1,*], G. S. Hassan[2] and T. A. Nahool[1 +]

[1] Physics Department, Faculty of Science at Qena, South Valley University, Egypt

[2] Physics Department, Faculty of Science, Assuit University, Egypt

[*] Yasser.mostafa@sci.svu.edu.eg
[+] tarek.abdelwahab@sci.svu.edu.eg

## Abstract

In this paper, we calculate the mass spectra of heavy quarkonium by using matrix Numerov's method to make the predictions of F and G states for further experiments. The method gives a very reasonable result which is in a good agreement with other methods and with recently published theoretical data. From the yielded wave functions we calculate the root mean square radius $r_{ms}$ and β coefficients of heavy quarkonium.

## 1. Introduction

The theoretical studies of the heavy quarkonium system [1] and its applications to bottomonium [2] and charmonium [3] is one of the special interest because of its relies entirely on the first principles of quantum chromo dynamics( QCD). From the viewpoint of the "heavy quarkonium" spectra, we calculate the theoretical mass spectra via non-relativistic frame work [4], [5], [6] by using matrix Numerov's method [7], [8] in two levels F and G. Many studies investigate "heavy quarkonium" properties within the quark model [9-15]. An essential progress has been made in the theoretical investigation of the non-relativistic heavy quark dynamics. Our point of departure is to calculate the spectra of heavy quarkonium and the corresponding wave functions of $n\bar{n}$ states (n = c, b) for the predicted F and G states. Actually, we don't have any experimental data of F and G states. Therefore, we compare the present theoretical predictions with published theoretical data from [16], [17] and [18]. Moreover, the heavy-meson wave functions determined in this work can be employed to make predictions of other properties. Besides, the main motivation is to calculate the root mean square radius $r_{ms}$ of different states for bottomonium and the numerical values of β coefficient [19], which can be used to calculate the decay widths [20], and differential cross sections [21] for quarkonium states. In this work, we consider the mass spectra and some properties of heavy quarkonium systems in the non-relativistic quark model using matrix Numerov's method. In section 2, we review the main formalism of the matrix Numerov's method used in our analysis and the used model. Besides, the analytical formula of the wave functions. Some characteristics properties of bottomonium mesons are introduced in section 3. After that, numerical results and discussion are given. Finally in the last section, we summarize our main results and conclusions.

## 2. Theoretical Basis

### 2.1. Matrix Numerov's Method

The Matrix Numerov's method is a method that provides us an approximate solution of non-relativistic Schrödinger equation of the form:

$$\psi''(x) = f(x)\psi(x) \qquad (1)$$

For the time-independent 1D Schrödinger equation, we have:

$$\frac{-\hbar^2}{2m} A_{N,N} \psi_i + B_{N,N} V_N \psi_i = E_i B_{N,N} \psi_i \qquad (2)$$

Where i run from 1 to N and define matrices

$$A_{N,N} = \frac{(I_{-1} - 2I_0 + I_1)}{d^2}, \; B_{N,N} = \frac{(I_{-1} + 10I_0 + I_1)}{12}, V_N = \text{diag}(\ldots, V_{i-1}, V_i, V_{i+1})$$

Where $I_{-1}$, $I_0$ and $I_1$ represent sub-, main-, and up- diagonal unit matrices respectively

For the 3D radial Schrödinger equation, the last equation could be written as:

$$\frac{-\hbar^2}{2\mu} A_{N,N} B_{N,N}^{-1} \psi_i + [V_N(r) + \frac{l(l+1)}{r^2}]\psi_i = E_i \psi_i \qquad (3)$$

Suggesting that this particle is compound and it consists of two smaller particles (meson consist of two quarks), then the reduced mass in the non-relativistic model can be identified as

$$\mu = \frac{m_q m_{\bar{q}}}{m_q + m_{\bar{q}}}$$

Where $m_q = m_{\bar{q}}$ is the mass of quark and anti-quark for Quarkonium system.

In natural units $\hbar = c = 1$ then, Eq. (3) could be written as:

$$\frac{-1}{2m} A_{N,N} B_{N,N}^{-1} \psi_i + [V_N(r) + \frac{l(l+1)}{r^2}]\psi_i = E_i \psi_i \qquad (4)$$

The first term is the Matrix Numerov's representation of the kinetic energy operator and the second is the Matrix Numerov's representation of the potential energy operator.

### 2.2. Non-relativistic Potential Quark Model

The non-relativistic quark model has been highly successful not only for heavy quarkonium $c\bar{c}$ and $b\bar{b}$ states, but also for states involving light quarks. As a minimal model of heavy quarkonium system, we use a non relativistic potential model, with wave functions determined by the Schrödinger equation. We use the more realistic model of the potential which is Coulomb plus linear plushyperfine interaction model [22], [23]

$$V(r) = -\frac{4\alpha_s}{3} + br + \frac{32\pi\alpha_s}{9m_q^2}\delta(r)S_q S_{\bar{q}} + \frac{l(l+1)}{2\mu r^2} \qquad (5)$$

Where

$$\delta(r) = (\frac{\sigma}{\sqrt{\pi}})^3 e^{-\sigma^2 r^2} \qquad (6)$$

And s is the total spin quantum number of the meson [24]

$$\text{and} \quad S_q S_{\bar{q}} = \frac{s(s+1)}{2} - \frac{3}{4} \qquad (7)$$

The hyperfine term is spin-dependent term which makes it possible to distinguish between mesons which have different spins. By including the tensor operator, the potential of the q$\bar{q}$ system has the following form for the bottomonium:

$$V_N(r) = \frac{l(l+1)}{2\mu r^2} - \frac{4\alpha_s}{3} + br + \frac{32\pi\alpha_s}{9m_b^2}\delta(r)S_b S_{\bar{b}} + \frac{1}{m_b^2}\left[\left(\frac{2\alpha_s}{r^3} - \frac{b}{2r}\right)\vec{L}.\vec{S} + \frac{4\alpha_s}{r^3}T\right] \quad (8)$$

## 2.3. Wave functions of bottomonium mesons

Quarkonium system can be described by the wave function of the bound quark-antiquark state which satisfies the Schrödinger equation (SE) by using the potential given in Equation (8). Radial Schrödinger equation, $\psi(r) = r\,R(r)$, is written (in natural units) as:

$$\psi(r) = 2\mu\,(E - V(r))\psi(r) = 0 \quad (9)$$

Where R(r) is the radial wave function, r is the inter quark distance, E is the sum of kinetic and potential of quark-antiquark system, and V (r) and μ are defined above through Equation (8). The matrix Numerove's method is used to solve Eq. (8) to get spectra of bottomonium $b\bar{b}$ as an example of heavy quarkonium mesons, the detailed of this method could be found in Ref. [7]. In the following sections, we employ that method to obtain the wave functions of bottomonium which in turn are used to determine some properties of bottomonium.

## 3. Basic Properties of Bottomonium Meson

### 3.1. Bottomonium root mean square radius $r_{ms}$

Bottomonium root mean square radius $r_{ms}$ is one of basic properties of bottomonium. If the distance between the quark and anti-quark in bottomonium is *r* fm it may be regarded that bottomonium has radius *r*/2 fm where *r* is the distance from the point quark to anti-quark. $r_{ms}$ can be derived from the meson wave function and may be written as [25].

$$r^2{}_{ms} = \int_0^\infty \{\psi^2(r)r^2 dr\} \quad (10)$$

### 3.2. β Coefficient

The meson wave function is characterized by a momentum width parameter β that is related to the root mean square quark-antiquark separation $r_{ms}$ of the meson by [26].

$$\beta = \sqrt{2(n-1) + (L) + \frac{3}{2}}\,\frac{1}{r_{ms}} \quad (11)$$

where n is the principal quantum number and L is the sub- atomic energy level number. The parameter β is typically taken as a parameter of the model. However, since we are seeking for describing the decay of heavy quark states, it is preferable to reproduce β coefficient of the quark model states. These values of β are obtained for the first time. So, we suggest using it to calculate the decay width of heavy quarkonium system.

## 4. Numerical results

### 4.1. Theoretical Predictions of Bottomonium spectrum

A non relativistic potential model is used to study bottomonium meson spectra by using the Numerov's Method. We predict the masses of the twelve $b\bar{b}$ states shown in **Table 1**, where we compared the present theoretical predictions with those from [16], [17] and [18]. **Figure 1**, **Figure 2** and **Figure 3**, illustrate the ratio between the obtained theoretical predictions of bottomonium spectra and those from [16], [17] and [18] respectively. It is seen that the ratio converges to one particularly in **Figure 1**. This means that the yielded results are in a good agreement with recently published predictions.

### 4.2. Theoretical Calculations of Some Properties of Bottomonium

The eigenvalues and the corresponding wave functions are found by using the same method. Then we normalized the wave functions and we calculate the root mean square radius of bottomonium mesons by using Equation (10). Moreover, we obtain computational values of β coefficient by using Equation (11). The normalized radial wave functions for bottomonium mesons are graphically represented in **Figure 4** and **Figure 5** respectively. For bottomonium mesons, our calculated masses and root mean square radius are reported in **Table 2** in case of F and G States respectively. The β values could be used to calculate the decay constants [27], decay widths [27], and differential cross sections [28] for quarkonium states with high accuracy as we used complicated potential model. The predictions about these quantities are also reported in **Table 2** for bottomonium F and G States respectively.

### 4.3. Figures and Tables

Table 1. Theoretical predictions of bottomonium mass spectra in [Gev].

| State | The theoretical bottomonium mass spectrums ||||  Ratio [our]/[18] | Ratio1 [our]/[16] | Ratio2 [our]/[17] |
|---|---|---|---|---|---|---|---|
| | *Theoretical masses in Gev [ours]* | *Theoretical masses in [GeV] [17]* | *Theoretical masses in [GeV] [16]* | *Theoretical masses in [GeV] [18]* | | | |
| $1^3F_4$ | 10.345 | 10.359 | 10.337 | 10.345 | 1 | 0.99864 | 1.0008 |
| $2^3F_4$ | 10.592 | 10.617 | 10.597 | 10.591 | 1.00009 | 0.99764 | 0.9995 |
| $1^3F_3$ | 10.344 | 10.355 | 10.340 | 10.344 | 1 | 0.99893 | 1.0004 |
| $2^3F_3$ | 10.591 | 10.613 | 10.599 | 10.588 | 1.0003 | 0.99792 | 0.9993 |
| $1^3F_2$ | 10.342 | 10.351 | 10.341 | 10.342 | 1 | 0.99913 | 1.0001 |
| $2^3F_2$ | 10.589 | 10.609 | 10.599 | 10.591 | 1.00009 | 0.99811 | 0.9991 |
| $1^1F_3$ | 10.344 | 10.355 | 10.339 | 10.344 | 1 | 0.99893 | 1.0005 |
| $2^1F_3$ | 10.591 | 10.613 | 10.598 | 10.591 | 1 | 0.99792 | 0.9993 |
| $1^3G_5$ | 10.5 | | | 10.501 | 0.99990 | | |
| $1^3G_4$ | 10.501 | | | 10.501 | 1 | | |
| $1^3G_3$ | 10.501 | | | 10.5 | 1.00009 | | |
| $1^1G_4$ | 10.5 | | | 10.5 | 1 | | |

Table 2. Theoretical predictions of the bottomonium $r_{ms}$ and the corresponding values of β coefficent.

| The theoretical prediction of $r_{ms}$ and the values of β coefficient | | |
|---|---|---|
| state | $r_{ms}$ [fm] | β values |
| $1^3F_4$ | 3.46755 | 0.611763 |
| $2^3F_4$ | 4.58349 | 0.556238 |
| $1^3F_3$ | 3.44305 | 0.616117 |
| $2^3F_3$ | 4.55974 | 0.559135 |
| $1^3F_2$ | 3.42368 | 0.619602 |
| $2^3F_2$ | 4.54079 | 0.561469 |
| $1^1F_3$ | 3.44926 | 0.615007 |
| $2^1F_3$ | 4.56573 | 0.558401 |
| $1^3G_5$ | 4.03734 | 0.58088 |
| $1^3G_4$ | 4.02281 | 0.582977 |
| $1^3G_3$ | 4.01083 | 0.584718 |
| $1^1G_4$ | 4.02576 | 0.582551 |

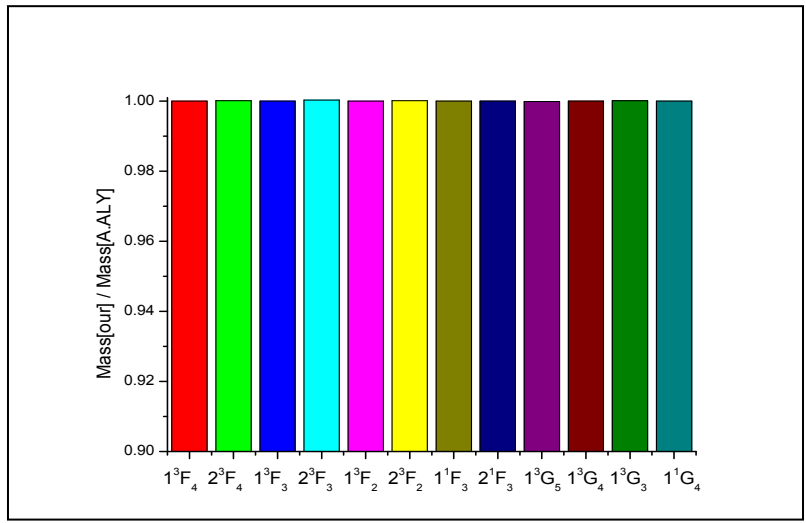

Figure 1: Ratios of the obtained theoretical masses of bottomonium F and G states to the theoretical data in [18].

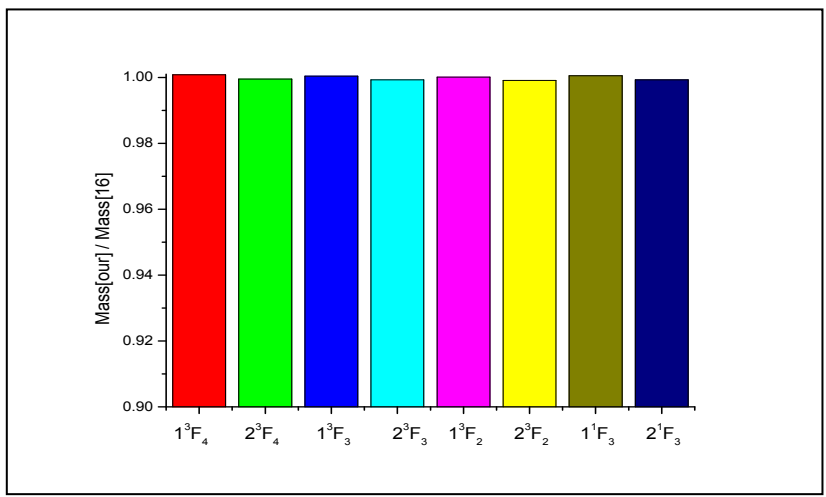

Figure 2. Ratios of the obtained theoretical masses of bottomonium G states to the theoretical data in [16].

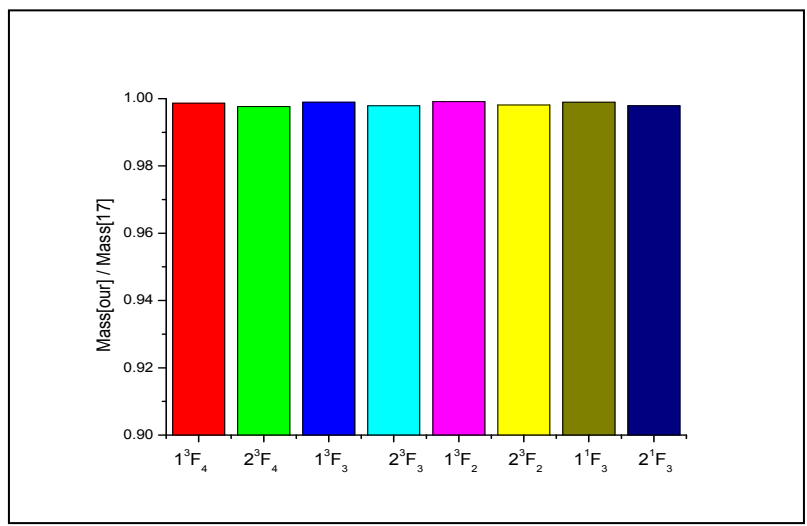

Figure 3. Ratios of the obtained theoretical masses of bottomonium G states to the theoretical data in [17]

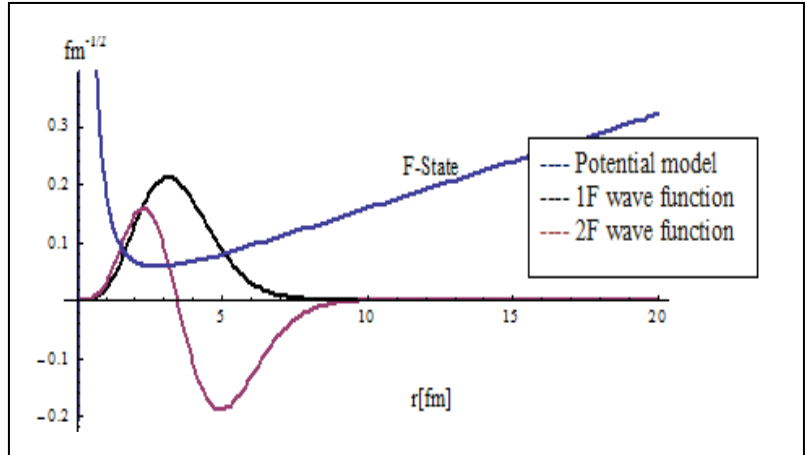

Figure 4. Bottomonium F-states reduced radial wave functions plotted together with the used potential.

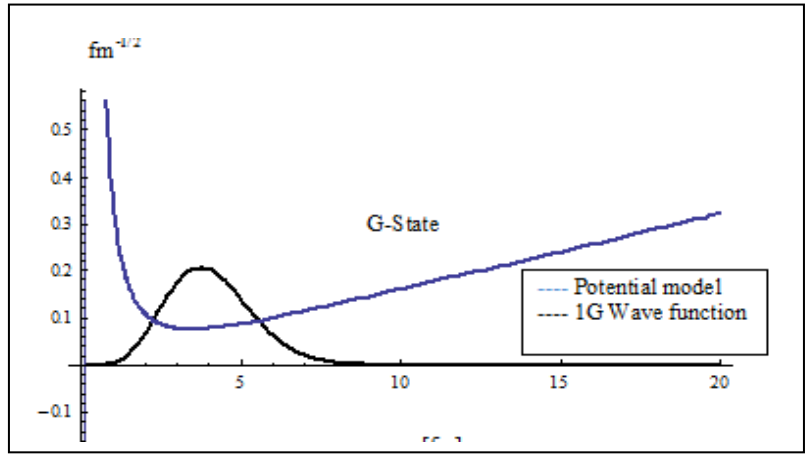

Figure 5. Bottomonium G-state reduced radial wave functions plotted together with the used potential.

## 5. SUMMARY AND CONCLUSION

In this paper, the mass spectra of bottomonium as an example of heavy quarkonium were studied within the framework of non-relativistic frame work. Eigenvalues and Eigenfunctions were obtained numerically for bottomonium meson using matrix Numerov's method. The predictions from our method are found to be in good agreement with the available theoretical results. Our calculated masses are reported in Table 1 along with the theoretical predictions of the other works. We observed that our results are in good agreement with existing theoretically predicted values, which shows the validity of the used method. Besides, there are an additional properties have been studied in this work since we used the matrix Numerov's method to obtain the radial wave functions of bottomonium meson to calculate the bottomonium $b\bar{b}$ root mean square $r_{ms}$ and β coefficient. As a remarkable result, we can point out that it is recommended to use the obtained values of $r_{ms}$ and β coefficient to calculate the decay widths and differential cross sections for bottomonium system. Moreover, the matrix Numerov's method [7] is tested again to obtain some mesons properties. Then, the method could be safely used to solve SE. Our calculated root mean square radius and the values of β coefficient are reported in Table 2. As a side result, these theoretical results are expected to give some hints to the forthcoming experiments. Eventually, we may notice that the calculated values of $r_{ms}$ and other parameters are the newer outputs where we didn't find others for comparison. So, we are looking forward to take these data in consideration by other experimental or theoretical researchers.